\documentclass[showpacs,prl,twocolumn,superscriptaddress]{revtex4}

\usepackage{graphicx}

\newcommand{\be}{\begin{equation}}
\newcommand{\ee}{\end{equation}}
\newcommand{\ba}{\begin{eqnarray}}
\newcommand{\ea}{\end{eqnarray}}
\newcommand{\bd}{\begin{description}}
\newcommand{\ed}{\end{description}}

\renewcommand{\iota}{{\bf 1}}
\def\rellow#1#2{Mathrel{Mathop{\kern 0pt #1}\limits_{#2}}}

\begin{document}

\title{(A note on) self-similarity in granular media}
\author{Lautaro Vergara}
\email{lvergara@lauca.usach.cl}
\affiliation{Departamento de F\'{\i}sica, Universidad de Santiago de Chile, Casilla 307,
Santiago 2, Chile }
\date{today}

\begin{abstract}
It is shown that the phenomenon of self-similarity appears in
granular media, with an intergrain potential $V \propto
\delta^{p+1}$, $p > 1$, where $\delta$ is the overlap between the
grains. Although this fact can be traced back in the literature,
this has not been put explicitly.
\end{abstract}

 \maketitle


Granular materials are nonlinear media which are of practical
interest for many applications. They are of interest also from the
theoretical point of view: in one-dimensional systems it has been
shown by theoretical \cite{Nesterenko,Wattis} as well as
experimental \cite{Coste} research, the existence of localised
travelling wave solutions. In practice, these solitary waves are
generated by impacting one end of a chain of grains. Despite the
large amount of recent work on the subject
\cite{Sen1,Hinch,Manciu,SenMan,Coste,Hong,SenProc,Naka,Rosas,Sen2,Nesterenko2,Nest,Nest3,Melo,Vergara},
the physics of granular media remains a challenge.

 Scaling laws have wide applications in science and engineering
\cite{Sedov,Barenb,Golden}. They give evidence of the property of
self-similarity of phenomena, that is of the fact that they
reproduce themselves after a rescaling of some variables and/or
parameters. Self-similarity has been used in the past in order to
transform systems of partial differential (PDEs) equations into
systems of ordinary differential equations, with the hope that the
initial problem will get easier to solve. This is one of the
standard methods for obtaining exact solutions of PDEs. Nowadays,
the search for exact solutions is focused to understand the
mathematical structure of the solutions and, hence, to get a
deeper understanding of the physical phenomena described by them.

In this Letter, it is shown that the phenomenon of self-similarity
appears in one-dimensional granular media with intergrain
potentials $V \propto \delta^{p+1}$, $p > 1$, where $\delta$ is
the overlap between the grains, in case there is no
pre-compression.

To illustrate the phenomenon of self-similarity in granular media,
lets us consider the scattering of solitary waves in a
one-dimensional linear chain interacting with a potential as the
one mentioned previously and with the form of a granular container
\cite{HA,H}, as shown in Fig. 1.

We shall assume that the granular container has a total of $M$
beads. There are two set of beads with $N_{1}$ beads located on
the lhs, $N_{2}$ on the rhs, both sets have beads with radii $a$.
Between them there are $L$ beads with radii $b$ ($a>b$).

Let $x_{i}(t)$ be the displacement of the center of the
$i$-th grain from its initial equilibrium position, and assume that the $i$%
-th grain, of mass $m_{i}$, has neighbors of different radii
and/or mechanical properties. Then, in absence of load and in a
frictionless medium, the equation of motion for the $i$-th sphere
reads

\begin{equation}
m_{i}\frac{d^{2}x_{i}}{dt^{2}}%
=k_{1}(x_{i-1}-x_{i})^{p}-k_{2}(x_{i}-x_{i+1})^{p}, \label{uno}
\end{equation}%
where it is understood that the brackets take the argument value
if they are positive and zero otherwise, ensuring that the spheres
interact only when in contact.

\begin{figure}[tbp]
\centering 
\hspace{-.5 cm} \includegraphics[width=.35\textwidth]{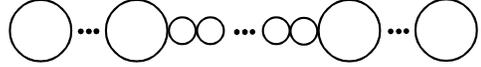} \vspace{%
-.2cm} \caption{{Schematic granular container used in the
calculations.}} \label{panel}
\end{figure}

In order to have realistic results, we shall assume that the
system consists of stainless-steel beads (see \cite{Coste} for
their properties), with radii $a=4$ mm and $b=2$ mm. The number of
beads is $N_{1}=20$, $N_{2}=20$ and $L=100$. We also choose
$\beta=10^{-5}$ m, $2.36 \,\times 10^{-5}$ kg and $\alpha = 1.0102
\,\times 10^{-3}$ s as units of distance, mass and time,
respectively. Through out the paper we assume that initially all
beads are at rest, that is,

\begin{eqnarray*}
u_{i}(0) &=&0,\text{ \ }i=1,\ldots ,M\\
\dot{u}_{i}(0) &=&0,\text{ \ }i=2,\ldots ,M \, ,  \label{init}
\end{eqnarray*}
except for the first bead at the left side of the chain. This bead
is supposed to have a nonzero value of velocity
$\dot{u}_{1}(0)=v_{0}$ in order to generate the soliton-like
perturbation in the chain. A simple analysis of the behavior of
the granular system under rescalings of the impact velocity from
$v_{0}$ to $\lambda \,v_{0}$, shows that the solutions of the set
of ODEs is self-similar. That is, it is found that

\begin{equation}
X(t;\lambda \, v_{0})=\lambda^{2/(p+1)} \,
X(\lambda^{(p-1)/(p+1)}\, t;v_{0}), \label{dos}
\end{equation}
where $X(t;v_{0})=\{x_i(t;v_{0}), \,\, \text{for all} \,\,
i=1,...,M\}$. This corresponds to the so called-power
self-similarity of the second kind, that is, the one that is
defined from the dynamics \cite{Barenb}. Of course, this
one-parameter transformation posses the group property
$T_{\lambda_1}\cdot T_{\lambda_2}=T_{\lambda_1 \,\lambda_2}$,
where $T_{\lambda} \, X(t;v_{0}) = X(t;\lambda \, v_{0})$. There
is also a corresponding equation for velocities.

The system is studied numerically by using an explicit Runge-Kutta
method of 5th order based on the Dormand-Prince coefficients, with
local extrapolation. As step size controller we have used the
proportional-integral step control, which gives a smooth step size
sequence.

In Figure 2 we show both, the solution for bead 28 corresponding
to an impact velocity $v_{0}=0.2$ m/s and the one with a scaled
impact velocity $\lambda \, v_{0}$, with $\lambda=6$.

It is worth to mention that Chatterjee \cite{Chatterjee}, while
studying a system of $N$ identical particles with an intergrain
potential $V \propto \delta^{p+1}$, noticed that if a function
$\tilde X(t)$ satisfies the corresponding equations of motion,
then so does the function $\alpha^{2/(p-1)} \, \tilde X(\alpha \,
t)$, for any positive (arbitrarily chosen) number $\alpha$. It
seems that he didn't noticed the self-similarity of solutions.
There is of course a mapping that connect them: by choosing
$\alpha=\lambda^{(p-1)/(p+1)}$ and recognizing that $\lambda$
corresponds to the scale factor of the initial impact velocity our
result is recovered.

Also, it is important to stress that this self-similarity
phenomenon is tacitly present in many works in the subject,
starting from the works of Nesterenko \cite{Nesterenko}, through
the work of Hinch and Saint-Jean \cite{Hinch}, till the work of
Rosas and Lindenberg \cite{Rosas}, where dimensionless variables
where used.

As stressed by Shirkov \cite{S}, the symmetry found here is not a
symmetry of the physical system or the equations of the problem at
hand, but a symmetry of a solution considered as a function of the
relevant physical variables and suitable initial conditions. This
kind of symmetry can be related to the invariance of a physical
quantity described by this solution with respect to the way in
which the initial conditions are imposed.

To end up it is important to stress also that the self-similarity
found in this work only appears in granular systems without
pre-compression. In fact, if we allow the system to be
pre-compressed (in this case we include a wall at the right side
of the chain) the eqs. of motion read

\begin{equation}
m_{i}\frac{d^{2}x_{i}}{dt^{2}}%
=k_{1}(\l_0-(x_{i}-x_{i-1}))^{p}-k_{2}(\l_0 -(x_{i+1}-x_{i}))^{p},
\label{uno}
\end{equation}%
and for the same initial conditions we found the result shown in
Fig. 3, for $\l_0=10^{-4}$. It is clear then that self-similarity
is lost when the system is loaded. Of course, the breaking of
self-similarity is less noticeable for very low loading. This is
related to the fact that, in dimensionless variables, the
self-similarity transformation appears via scale transformations
(with parameter $v_{0}$) of time and space coordinates.

\begin{figure}[tbp]
\centering 
\hspace{-.5 cm} \includegraphics[width=.37\textwidth]{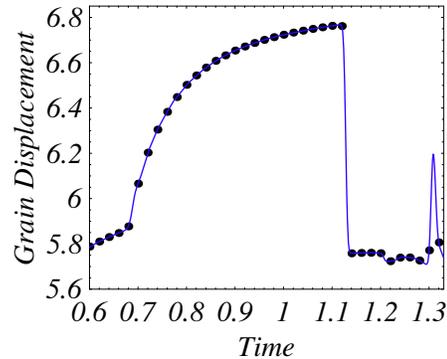} \vspace{%
-.2cm} \caption{{Displacement of bead 28 as a function of time (in
program units)for two different impact velocities $v_{0}=0.2$ m/s
(dots) and $v_{0}=1.6$ m/s (full line). }} \label{panel}
\end{figure}

\begin{figure}[h]
\centering 
\hspace{-.5 cm} \includegraphics[width=.37\textwidth]{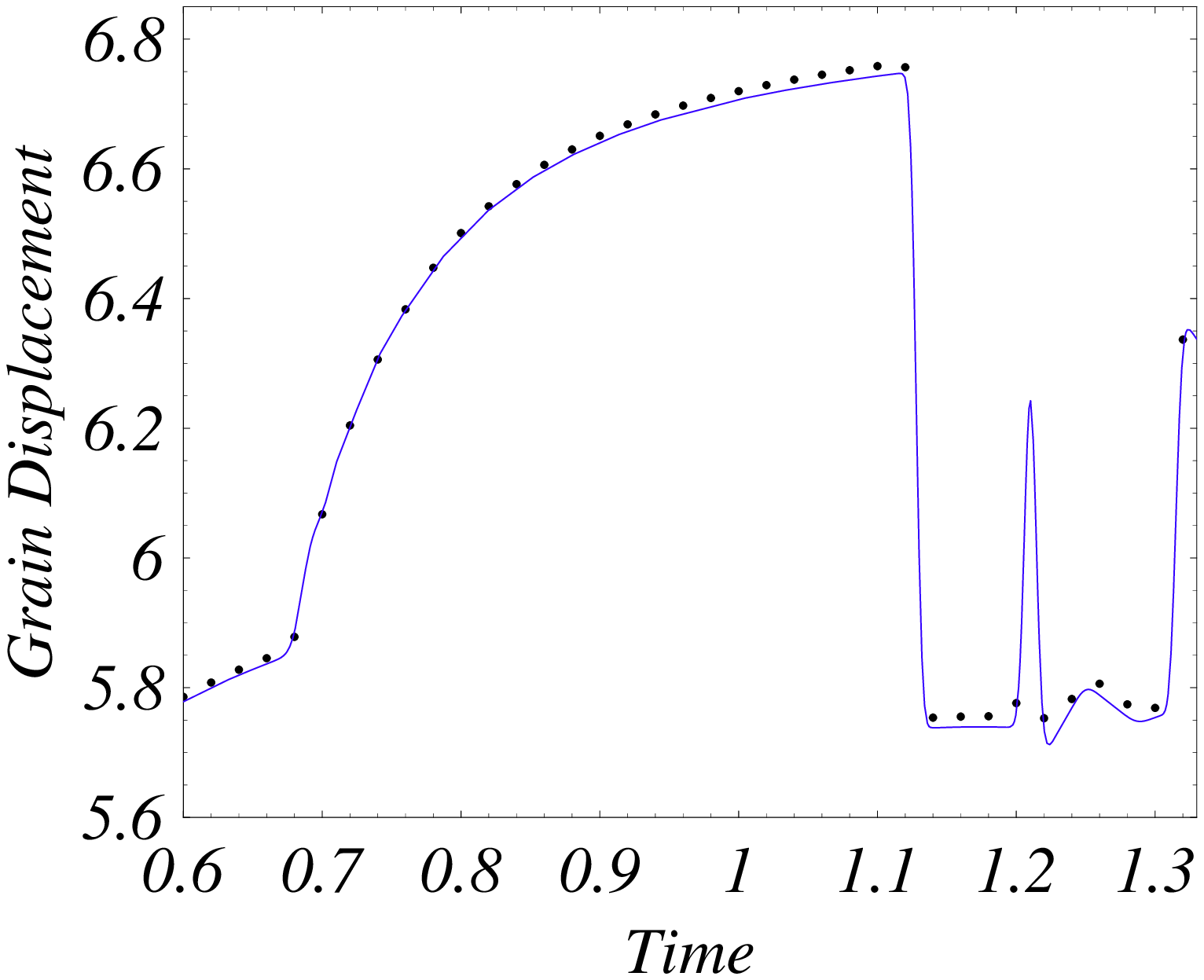} \vspace{%
-.2cm} \caption{{Displacement of bead 28 as a function of time (in
program units) for two different impact velocities $v_{0}=0.2$ m/s
(dots) and $v_{0}=1.6$ m/s (full line), in case of a
pre-compressed system. }} \label{panel}
\end{figure}



I want to acknowledge fruitful discussions with Dr. Ra\'ul
Labb\'e. I also acknowledge useful comments of Prof. Vitali F.
Nesterenko. This work was partially supported by DICYT-USACH No.
04-0531VC.


\begin{thebibliography}{99}
\bibitem{Nesterenko} V.F. Nesterenko, Zh. Prik. Mekh. Tekh. Fiz. \textbf{5}
(1983) 733; A.N. Lazaridi and V.F. Nesterenko, J. of Appl. Mech.
Tech. Phys. \textbf{26}, 405 (1985).

\bibitem{Wattis} G. Friesecke and J.A.D. Wattis, Commun. Math. Phys. \textbf{%
161} (1994) 391

\bibitem{Coste} C. Coste, E. Falcon, and S. Fauve, Phys Rev. E \textbf{56}
(1997) 6104

\bibitem{Sen1} S. Sen and R.S. Sinkovits, Phys Rev E \textbf{54} (1996) 6857

\bibitem{Hinch} E.J. Hinch and S. Saint-Jean, Proc. R. Soc. London, Ser. A
\textbf{455} (1999) 3201

\bibitem{Hong} J. Hong and A. Xu, Phys. Rev. E \textbf{63} (2001) 061310

\bibitem{Manciu} M. Manciu, S. Sen and A.J. Hurd, Physica D \textbf{157}
(2001) 226

\bibitem{SenMan} S. Sen and M. Manciu, Phys. Rev. E \textbf{64} (2001) 056605

\bibitem{SenProc} S. Sen \textit{et al.}, in \textit{Modern Challenges in
Statistical Mechanics: Patterns, Noise and the Interplay of Nonlinearity and
Clomplexity}, edited by V. M. Kenkre and K. Lindenberg, AIP Conference
Proceedings \textbf{658} 357 (2003) 357

\bibitem{Naka} M. Nakagawa, J. H. Agui, D. T. Wu and D. V. Extramiana,
Granular Matter \textbf{4} (2003) 167

\bibitem{Rosas} A. Rosas and K. Lindenberg, Phys. Rev. E \textbf{69} (2004)
037601


\bibitem{Sen2} \textit{The Granular State}, S. Sen and M.L. Hunt (Eds.),
Mater. Res. Soc. Symp. Proc. No. 627, Material Research Society, Pittsburg,
2001

\bibitem{Nesterenko2} \textit{Dynamics of Heterogeneous Materials} V.F.
Nesterenko, Springer-Verlag New York, 2001; V.F. Nesterenko, A.N.
Lazaridi and E.B. Sibiryakov, Jour. Aplied Mech. Tech. Phys., 36
(1995) 166.

\bibitem{Nest}  V.F. Nesterenko, C. Daraio, E.B. Herbold and S. Jin, Rev. Lett \textbf{95}, 158702 (2005)


\bibitem{Nest3}  C. Daraio, V.F. Nesterenko, E.B. Herbold and S. Jin, Phys. Rev. Lett. \textbf{96}, 058002
(2005).

\bibitem{Melo} S. Job, F. Melo, A. Sokolow and S. Sen, Phys. Rev. Lett \textbf{94}, 178002 (2005)

\bibitem{Vergara} L. Vergara, Phys. Rev. Lett \textbf{95}, 108002 (2005),
cond-mat/0503457


\bibitem{Sedov} L.I. Sedov, \textit{Similarity and Dimensional Methods in Mechanics},
(Academic Press, New York, 1959)

\bibitem{Barenb} G. I. Barenblatt, Scaling, Self-Similarity and Intermediate
Asymptotics (Cambridge University Press, Cambridge, UK, 1996)

\bibitem{Golden} N. Goldenfeld, Lectures on Phase Transitions and the Renormalization Group
(Addison-Wesley, Reading, Massachusetts, 1992)

\bibitem{HA} J. Hong and A. Xu, Appl. Phys. Lett., 81, 4868 (2002)

\bibitem{H} J. Hong, Phys. Rev. Lett. 94, 108001 (2005)


\bibitem{Chatterjee} A. Chatterjee, Phys. Rev. E \textbf{59}, 5912 (1999)


\bibitem{S}  D.~V.~Shirkov, Sov. Phys. Doklady \textbf{27} 197
(1982); Theor. Math. Phys. \textbf{60} 778 (1984); Intern. J. Mod.
Physics \textbf{A3} 1321 (1988)



\end{thebibliography}
\end{document}